\documentstyle[12pt]{article}
\makeatletter
\newcommand{\singlespacing}{\let\CS=\@currsize\renewcommand{\baselinestretch}
{1.0}\tiny\CS}
\newcommand{\doublespacing}{\let\CS=\@currsize\renewcommand{\baselinestretch}
{1.5}\tiny\CS}

\begin{document}
\begin{center}
{\Large {\bf Optimal Universal Disentangling Machine for Two Qubit Quantum States}}

\vspace{0.6cm}

{\bf Sibasish Ghosh}$^a$~\footnote{e-mail: res9603@isical.ac.in}, 
{\bf Somshubhro Bandyopadhyay}$^b$~\footnote{e-mail: dhom@bosemain.boseinst.ernet.in}, 
{\bf Anirban Roy}$^a$~\footnote{e-mail: res9708@isical.ac.in},\\ {\bf Debasis
Sarkar}$^c$~\footnote{Present address: Department of Mathematics, University of
Burdwan, Burdwan, West Bengal, India.} and {\bf Guruprasad Kar}$^a$ 
\end{center}

{\noindent $^a$ {\it Physics and Applied Mathematics Unit, Indian Statistical
Institute, 203 B.T. Road, Calcutta -700035, India.}}

{\noindent $^b$ {\it Department of Physics, Bose Institute, 93/1 A.P.C. Road,
Calcutta -700009, India.}}

{\noindent $^c$ {\it Department of Mathematics, Barrackpore Rastraguru Surendranath
College, 85 Middle Road, Barrackpore, North 24 Parganas, West Bengal, India.}}

\vspace{0.7cm}
\begin{center}
{\bf Abstract}

We derive the optimal curve satisfied by the reduction factors, in the case of universal disentangling
machine which uses only local operations. Impossibility of constructing a
better disentangling machine, by using non-local operations, is discussed.
\end{center}

\vspace{1.5cm}

\newpage
\section{Introduction}

Disentanglement is the process that transforms
a state of two (or more) subsystems into an unentangled state (in
general, a mixture of product states) such that the reduced density
matrices of each of the subsystems are unaffaected.

Let $\rho ^{ent}$ be any entangled state of two qubits 1 and 2; and let
$\rho _1$, $\rho _2$ be the reduced density matrices of 1 and 2 respectively.
Then the operation of any disentangling machine (DM) is defined as 
$$\rho ^{ent}~~~ \frac{\rm DM}{}\!\!\!\rightarrow ~~\rho ^{disent}$$  
together with 
$$\rho _i = {\rm Tr}_{j} (\rho ^{ent}) = {\rm Tr}_{j} (\rho ^{disent}),~ i
\neq j;~ i, j = 1, 2$$
for all $\rho ^{ent}$. 
This kind of ideal universal disentangling machine does not exist \cite{mor}, \cite{terno}.

So the next question is whether there exists a disentangling machine which
disentangles entangled state, and for which
$${\rm Tr}_{j} (\rho ^{disent}) = \eta _i \rho _{i} + \left( \frac{1 - \eta _i}{2} \right) I,~ i
\neq j;~ i, j = 1, 2$$
where $\eta _i$ ($0 < \eta _i < 1$ for $i = 1, 2$) is independent of $\rho
^{ent}$ \cite{bruss}. Recently it has been shown that \cite{som} this kind of machine
exists, by using local cloning operations, where the input states are all pure
entangled states. Reference \cite{som} considered two
cases, (1) $\eta _1 = 1$ (or $\eta _2 = 1$), {\it i.e.}, using only one local
cloning machine, (2) $\eta _1 = \eta _2$ ($= \eta$, say), {\it i.e.}, which uses two local
cloning machines with same fidelity. For the case (1), the maximum value of
$\eta _2$ (or $\eta _1$) is $1/3$. In the case (2), the maximum attainable
value of $\eta$ is $1/{\sqrt{3}}$. 
In the present paper, we
want to find out the optimum values of $\eta _1$ and $\eta _2$, or, in other
words, the optimal curve (if it exists) satisfied by $\eta _1$ and $\eta _2$
({\it i.e.}, reduction factors corresponding to the optimal
disentangling machine), by using most
general (asymmetric) local operations. Surprisingly, we got the same upper bounds on $\eta$ as has
been found in \cite{som}, in the corresponding cases (1) and (2). We have also
obtained the optimal curve in the most general case, when asymmetric local
operations are used.   

In section 2, for simplicity, we first consider disentanglement process by
applying our disentangling machine on one of the subsystems. In section 3, we
deal with the symmetric case  where the
same disentangling machine is used locally on the two subsystems. Next, in
section 4, the most general disentangling machine is considered using asymmetric local
operations, where we discuss the disentanglement of mixed states. In section 5
we sum up our results and put some
arguments regarding nonlocal operations.
        
\section{Totally asymmetric optimal universal disentangling machine}

In this section, we shall consider how we can disentangle a two qubit pure entangled
state by local operation on any one qubit. Suppose we have two parties 
$x$ and $y$ sharing an entangled state of two qubits given by
\begin{equation}
\label{twoqubit}
\left| \psi \right\rangle = \alpha \left| 00\right\rangle _{xy} + \beta \left| 11\right\rangle _{xy},
\end{equation}
where $\alpha$ and $\beta$ are non-negative numbers with ${\alpha}^2 +
{\beta}^2 = 1$. The first qubit belongs to $x$ and the second belongs to $y$ as usual.
Now a universal transformation (a unitary operation) is applied to the qubit belonging to any one (say, 
$y$) of the two parties. This gives
rise to a composite system  ${\rho}_{xyM}$ consisting of the two qubits and a
(disentangling) machine $M$. Tracing out
on the machine states we get a
two qubit composite system which is disentangled ({\it i.e.}, separable) under
certain conditions.

Consider the following unitary transformation $U_j$ (associated with a machine
state $|M\rangle _j$) applied on one subsystem $j$ (where $j = x$ or $y$),
defined by
\begin{equation}
\label{U0}
U_j\left|0 \right\rangle _j\left|M \right\rangle _j = m_{0j} \left|0 \right\rangle
_j\left|M_0 \right\rangle _j + m_{1j} \left|1 \right\rangle _j\left|M_1
\right\rangle _j,
\end{equation}
\begin{equation}
\label{U1}
U_j\left|1 \right\rangle _j\left|M \right\rangle _j = \widetilde{m_{0j}} \left|0 \right\rangle
_j\left|\widetilde{M_0} \right\rangle _j + \widetilde{m_{1j}} \left|1
\right\rangle _j\left|\widetilde{M_1} \right\rangle _j,
\end{equation}
where $\left \{\left|M_0 \right\rangle _j, \left|M_1 \right\rangle _j, 
\left|\widetilde{M_0} \right\rangle _j, \left|\widetilde{M_1} \right\rangle _j
\right\}$ are four normalized  machine states, and (using unitarity) 
\begin{equation}
\label{unitari}
\left.
\begin{array}{lcr}
\left|m_{0j} \right|^2 + \left|m_{1j} \right|^2 &=& 1,\\
\left|\widetilde{m_{0j}} \right|^2 + \left|\widetilde{m_{1j}} \right|^2 &=& 1.
\end{array}
\right \}
\end{equation}
Using orthogonality, we have (from (2) and (3)),
\begin{equation}
\label{ortho}
{m_{0j}^* \widetilde{m_{0j}}} _j\!\left\langle M_0 | \widetilde{M_0}
\right\rangle_j + {m_{1j}^*
\widetilde{m_{1j}}} _j\!\left\langle M_1 | \widetilde{M_1} \right\rangle_j = 0.
\end{equation}  

Now applying this operation $U_j$ on an arbitrary one qubit state $|\phi
\rangle_j = a
|0\rangle_j + b |1\rangle_j$ (where $|a|^2 + |b|^2 = 1$), we get the
following composite state,
$$U_j \left|\phi \right\rangle_j \left|M \right\rangle_j = $$
$$a \left[ m_{0j} \left|0 \right\rangle_j \left|M_0 \right\rangle_j + m_{1j} \left|1
\right\rangle_j \left|M_1 \right\rangle_j \right] + b \left[
\widetilde{m_{0j}} \left|0 \right\rangle_j \left|\widetilde{M_0}
\right\rangle_j + \widetilde{m_{1j}} \left|1 \right\rangle_j
\left|\widetilde{M_1} \right\rangle_j \right]$$
\begin{equation}
\label{composite}
= a m_{0j} \left|0 \right\rangle_j \left|M_0 \right\rangle_j + a m_{1j}
\left|1 \right\rangle_j \left|M_1 \right\rangle_j + b \widetilde{m_{0j}}
\left|0 \right\rangle_j \left|\widetilde{M_0} \right\rangle_j + b
\widetilde{m_{1j}} \left|1 \right\rangle_j \left|\widetilde{M_1} \right\rangle_j. 
\end{equation}

We got (6) by applying the above unitary operation on any one qubit ($x$ or
$y$) pure state $|\phi \rangle_j$ ($j = x, y$), where
$\rho_j = |\phi {\rangle\!_j} {_j\!\langle} \phi| = \frac{1}{2} ({\bf 1} + \vec{s} .
\vec{\sigma})$ (with $|\vec{s}| = 1$). And now we demand that the reduced 
density matrix, after tracing out the machine states in the equation 
(\ref{composite}), is of the form  $\frac{1}{2} ({\bf 1}
+ \eta_j \vec{s} . \vec{\sigma})$ (where $0 < \eta_j \le 1$) for all $\vec{s}$
(isotropy) \cite{bruss}. Then the machine has
to satisfy the following equations :
\begin{equation}
\label{machine1}
\left.
\begin{array}{lcr}
{m_{0j} m_{1j}^*} _j\!\left\langle M_1 | M_0 \right\rangle_j &=& 0,\\
{\widetilde{m_{0j}} \widetilde{m_{1j}}^*} _j\!\left\langle \widetilde{M_1} |
\widetilde{M_0} \right\rangle_j &=& 0,\\
{m_{1j}^* \widetilde{m_{0j}}} _j\!\left\langle M_1 |
\widetilde{M_0} \right\rangle_j &=& 0.
\end{array}
\right \}
\end{equation}
\begin{equation}
\label{machine2}
\left.
\begin{array}{lcr}
{\rm Re} \left \{ {m_{0j}^* \widetilde{m_{0j}}} _j\!\left\langle M_0 |
\widetilde{M_0} \right\rangle_j \right \} &=& 0,\\
{\rm Re} \left \{ {m_{1j}^* \widetilde{m_{1j}}} _j\!\left\langle M_1 |
\widetilde{M_1} \right\rangle_j \right \} &=& 0.
\end{array}
\right \}
\end{equation}
\begin{equation}
\label{machine3}
\eta_j = {m_{0j} \widetilde{m_{1j}}^*} _j\!\left\langle \widetilde{M_1} |
M_0 \right\rangle_j.
\end{equation}
\begin{equation}
\label{machine4}
\left.
\begin{array}{lcr}
\left| m_{0j} \right| &=& \displaystyle{\left(\frac{1 + \eta _j}{2} \right)^{1/2}},\\
\left| m_{1j} \right| &=& \displaystyle{\left(\frac{1 - \eta _j}{2} \right)^{1/2}},\\
\left| \widetilde{m_{0j}} \right| &=& \displaystyle{\left(\frac{1 - \eta _j}{2} \right)^{1/2}},\\
\left| \widetilde{m_{1j}} \right| &=& \displaystyle{\left(\frac{1 + \eta _j}{2} \right)^{1/2}}.
\end{array}
\right \}
\end{equation}

Now we apply the above-mentioned operation on one of the two qubits (say on
$y$ in the state $|\psi\rangle$, in equation (\ref{twoqubit})\footnote{As the
system $x$ is unchanged, therefore, $\eta _x = 1$}). Then the state $\left| \psi \right\rangle = \alpha \left| 00\right\rangle_{xy} + \beta \left| 11\right\rangle_{xy}$ is transformed (after tracing out
the machine states, and applying all the above conditions {\it i.e.},
(\ref{unitari}), (\ref{ortho}), (\ref{machine1}) --
(\ref{machine4})) to the following density matrix :
$$D_{xy}^{TA} = {\rm Tr}_M \left[ {\rho}_{xyM} \right] = $$
\begin{equation}
\label{disent1}
\left[
\begin{array}{cccc}
\frac{{\alpha}^2 \left(1 + \eta_y \right)}{2} & 0 & -i \alpha \beta \Lambda_y & \alpha
\beta \eta_y\\
0 & \frac{{\alpha}^2 \left(1 - \eta_y \right)}{2} & 0 & i \alpha \beta \Lambda_y\\
i \alpha \beta \Lambda_y & 0 & \frac{{\beta}^2 \left(1 - \eta_y \right)}{2} & 0\\
\alpha \beta \eta_y & -i \alpha \beta \Lambda_y & 0 & \frac{{\beta}^2 \left(1 +
\eta_y \right)}{2}
\end{array}
\right]
\end{equation}
where $\Lambda_j = {\rm Im} \left \{ {m_{0j}^* \widetilde{m_{0j}}}
_j\!\left\langle M_0 | \widetilde{M_0} \right\rangle_j \right \}$ (for $j = x, y$), and the entries of
this matrix are arranged in accordance with the ordered basis $\{|00\rangle,
|01\rangle, |10\rangle, |11\rangle \}$ of the two qubits. 

Now we apply the Peres--Horodecki theorem to test the inseperability of 
$D_{xy}^{TA}$, which states that a density matrix $\rho$
of a bipartite system in ${C\!\!\!\!I}~ ^2 \otimes {C\!\!\!\!I}~ ^2$ is separable if and
only if the partial transposition of $\rho$ is positive semidefinite, {\it
i.e.}, each of the eigen values of the partial transposition of $\rho$ is
non-negative \cite{peres}, \cite{horodecki}, \cite{sanpera}.
It turns out that the state $|\psi\rangle$ will
be disentangled ({\it i.e.}, $D_{xy}^{TA}$ is separable) if the following
conditions are satisfied:
\begin{equation}
\label{TA}
\left.
\begin{array}{lcc}
1 - \eta_y^2 + 2 \alpha^2 \beta^2 \left(1 - \eta_y^2 - 4 \Lambda_y^2 \right) \ge 0,\\
\alpha^2 \beta^2 \left\{(1 + \eta_y)^2 (1 - 2 \eta_y) - 4 \Lambda_y^2 \right\} \ge 0,\\
\alpha^4 \beta^4 \left\{(1 - 3 \eta_y) (1 + \eta_y)^3 + 8 \Lambda_y^2 \left(2
\Lambda_y^2 - 1 + \eta_y^2 \right)\right\} \ge 0.
\end{array}
\right \}
\end{equation}
All the three conditions in (\ref{TA}) will be satisfied for all $\alpha \beta
\in [0, 1/2]$ (as we are seeking for universal disentangling machine), if
$\eta_y \le 1/3$.

Thus we see that {\it all pure states of two qubits $x$ and $y$ will be disentangled by
applying a disentangling machine locally on $y$, provided $\eta_y \le 1/3$}. 

Now that our requirement is also to have reduced density matrices
$D_{ad}^{(x)} = {\rm Tr}_y [D_{xy}^{TA}]$, $D_{ad}^{(y)} = {\rm Tr}_x [D_{xy}^{TA}]$
of the disentangled state $D_{xy}$ as close as possible to those of the entangled
one ({\it i.e.}, ${\rho}_{bd}^{(x)} = {\rm Tr}_y [|\psi \rangle \langle \psi|]$ and ${\rho}_{bd}^{(y)} = {\rm
Tr}_x [|\psi \rangle \langle \psi|]$ respectively), we note that the reduced density matrix of the subsystem $x$ is
unaltered whereas that of the subsystem $y$ is changed.\\ 
Let us summarise these results.

1. It is possible to disentangle any arbitrary bipartite entangled
state by applying local disentangling machine on one of its qubits provided the
reduction factor ($\eta$) of the isotropic machine is less than or equal to
1/3.

2. After disentanglement the reduced density matrices of the subsystems
are given by
$$D_{ad}^{(x)} = {\rho}_{bd}^{(x)}$$ 
$$D_{ad}^{(y)} = \eta {\rho}_{bd}^{(y)} + \left( \frac{1 - \eta}{2} \right) I$$ 
where ${\eta}_{max} = 1/3$.   

\section{Symmetric optimal universal disentangling machine}

In the previous section we have shown how to disentangle any pure state of two qubits 
by applying local operation on one of the qubits. In this section we 
apply the local unitary operation $U = U_x = U_y$\footnote{so here $\eta_x =
\eta_y \equiv \eta$ (say), $m_{ix} = m_{iy}$ ($i = 0, 1$), $\widetilde{m_{ix}} =
\widetilde{m_{iy}}$ ($i = 0, 1$), $\left| M_i
\right\rangle_x = \left| M_i \right\rangle_y$ ($i = 0, 1$), $\left| \widetilde{M_i}
\right\rangle_x = \left| \widetilde{M_i} \right\rangle_y$ ($i = 0, 1$),
$\Lambda_x = \Lambda_y \equiv \Lambda$ (say)}, defined by 
equations (\ref{U0}) and (\ref{U1}), on both the parties $x$ and $y$ (in the state $|\psi\rangle$, given
in equation (\ref{twoqubit})) separately. 

Each of the two parties now performs the same local unitary operation $U$ on their own
qubit, as described in the previous section. After this local operation, 
the reduced density matrix (tracing out the machine states) of the two parties $x$ and $y$ (applying all the 
constraints on the machine states, {\it i.e.}, conditions (\ref{unitari}), (\ref{ortho}), (\ref{machine1}) --
(\ref{machine4})) is given by,
$${\rho}_{xy} = $$
\begin{equation}
\label{disent2}
\left[
\begin{array}{cccc}
\frac{(1 - \eta)^2}{4} + {\alpha}^2 \eta - 2 \alpha \beta {\Lambda}^2 
& -i \alpha \beta \Lambda \eta & -i \alpha \beta \Lambda \eta & \alpha
\beta {\eta}^2\\
i \alpha \beta \Lambda \eta & \frac{1 - {\eta}^2}{4} + 2 \alpha \beta {\Lambda}^2 & 0 & i \alpha \beta \Lambda \eta\\
i \alpha \beta \Lambda \eta & 0 & \frac{1 - {\eta}^2}{4} + 2 \alpha \beta {\Lambda}^2 & i \alpha \beta \Lambda \eta\\
\alpha \beta {\eta}^2 & -i \alpha \beta \Lambda \eta & -i \alpha \beta \Lambda \eta & \frac{(1 - \eta)^2}{4} + {\beta}^2 \eta - 2 \alpha \beta {\Lambda}^2 
\end{array}
\right]_.
\end{equation}

It follows from the Peres-Horodecki theorem \cite{peres}, \cite{horodecki}, \cite{sanpera}, that
${\rho}_{xy}$ is separable ({\it i.e.}, the state $|\psi\rangle$ is
disentangled) if
\begin{equation}
\label{c1}
a_1 \left(2a_2 + a_3 \right) + a_2 \left(a_2 + 2 a_3 \right) - (\alpha
\beta \eta)^2 (4 {\Lambda}^2 + {\eta}^2) \ge 0,
\end{equation} 
$$a_1 a_2 \left(a_2 + 2a_3 \right) + a_2^2 a_3 - \left(a_1 + a_3 \right)
(\alpha \beta \eta)^2 (2{\Lambda}^2 + {\eta}^2)$$
\begin{equation}
\label{c2}
- 4a_2 (\alpha \beta \eta \Lambda)^2 - 4(\alpha \beta)^3 {\eta}^4 {\Lambda}^2
\ge 0,
\end{equation}

\newpage

\begin{equation}
\label{c3}
a_1 a_2^2 a_3 - 2a_2 \left(a_1 + a_3 \right) (\alpha \beta \eta \Lambda)^2 - a_1 a_3 (\alpha \beta)^2 {\eta}^4 - 2 \left(a_1 + a_3 \right) (\alpha \beta)^3
{\eta}^4 {\Lambda}^2 \ge 0, 
\end{equation}
where 
$$a_1 = \frac{(1 - \eta)^2}{4} + {\alpha}^2 \eta - 2 \alpha \beta {\Lambda}^2,$$
$$a_2 = \frac{1 - {\eta}^2}{4} + 2 \alpha \beta {\Lambda}^2,$$
$$a_3 = \frac{(1 - \eta)^2}{4} + {\beta}^2 \eta - 2 \alpha \beta {\Lambda}^2.$$ 

Now we shall consider the following two special cases, where in the first case,
we put constraint on the machine, and in the second, we take the original state
as a maximally entangled state.\\
{\noindent {\it Case I} : $\Lambda = 0$, {\it i.e.}, ${\rm Im} \left\{\left\langle M_0 |
\widetilde{M_0} \right\rangle \right\} = 0$ \cite{fn1}.}   
 
In this case, conditions (\ref{c1}) -- (\ref{c3}) will be reduced to the
following three conditions respectively:
\begin{equation}
\label{L0c1}
\frac{1 - \eta^2}{8} \{3 + \eta^2 + 8 (\alpha \beta \eta)^2\} \ge 0,
\end{equation}
\begin{equation}
\label{L0c2}
(1 - \eta^2)^2 + 8 (\alpha \beta \eta)^2 (1 - 2 \eta^2 - \eta^4) \ge 0,
\end{equation}
\begin{equation}
\label{L0c3}
\left\{\frac{(1 - \eta^2)^2}{16} + (\alpha \beta \eta)^2 \right\} \left\{\frac{(1 - \eta^2)^2}{16} - (\alpha \beta \eta^2)^2 \right\} \ge 0.
\end{equation}

It is clear from the above conditions (\ref{L0c1}) -- (\ref{L0c3}) that 
all bipartite pure entangled states ({\it i.e.}, for all $\alpha \beta \in [0, 1/2]$), 
will be disentangled, if the reduction factor ($\eta$) is
less than or equal to $1/{\sqrt{3}}$, and so the maximum value ${\eta}_{max}$
of $\eta$ is equal to $1/{\sqrt{3}}$.\footnote{Note the error in equation no. (17) of
\cite{buz}, and also in \cite{som}.}  

{\noindent {\it Case II} : $\alpha \beta = 1/2$ \cite{fn2}.}

Here, above conditions (\ref{c1}) -- (\ref{c3}) will be reduced to the
following three conditions respectively:
\begin{equation}
\label{halfc1}
\Lambda^4 \le \frac{3}{16} (1 - \eta^4),
\end{equation}
\begin{equation}
\label{halfc2}
\Lambda^4 \le \frac{1 - 3 \eta^4 - 2 \eta^6}{16},
\end{equation}
\begin{equation}
\label{halfc3}
(1 + \eta^2 + 4 \Lambda^2) (1 + \eta^2 - 4 \Lambda^2) (1 - 2 \eta^2 - 3 \eta^4
+ 8 \Lambda^2 \eta^2 - 16 \Lambda^4) \ge 0.
\end{equation}

As we have to maximize $\eta$, we have to check (\ref{halfc1}) --
(\ref{halfc3}) with all possible values of
$\Lambda$. For that reason, we take, in the above three conditions
(\ref{halfc1}) -- (\ref{halfc3}), the values of $\Lambda^2$, starting from 0.
Keeping all these in mind, it can be shown that all the conditions
(\ref{halfc1}) -- (\ref{halfc3}) will be satisfied for all $\alpha \beta \in
[0, 1/2]$, if the maximum value of $\eta$ is $1/{\sqrt{3}}$.

As mentioned in footnote 6, $\Lambda \equiv {\rm Im} \left\{m_0^*
\widetilde{m_0} \left\langle M_0 |
\widetilde{M_0} \right\rangle \right\} = \sqrt{(1 - {\eta}^2)/4}~~ \times$\\
${\rm Im} \left\{\left\langle M_0 |
\widetilde{M_0} \right\rangle \right\} = \sqrt{(1 - {\eta}^2)/4}~ \lambda$,
where $\lambda = {\rm Im} \left\{\left\langle M_0 |
\widetilde{M_0} \right\rangle \right\}$, so $\lambda \in [-1, 1]$.  
In the general situation, for an arbitrary (universal) disentangling
machine ({\it i.e.}, for arbitrary value of $\lambda \in [-1, 1]$), we have to test whether
the maximum value ${\eta}_{max}$ of the reduction factor can be made greater 
than $1/{\sqrt{3}}$. Now we note that the conditions given in
(\ref{c1}) to (\ref{c3}) are non-linear in ${\lambda}^2$, $\alpha \beta$ and
$\eta$, making it very difficult to get ${\eta}_{max}$ analytically from these
conditions. And so we proceed numerically. As we are concerned with universal
disentangling machines (different machines correspond to different values of
$\lambda$), therefore, we have to find out the maximum value $\eta _{max}
({\lambda}^2)$ among all possible values $\eta ({\lambda}^2)$ of the reduction factor 
for which all the states ({\it i.e.}, for all
the values of $\alpha \beta \in [0, 1/2]$) will be disentangled by a
disentangling machine corresponding to the given value of $\lambda$, so that
all the conditions (\ref{c1}) -- (\ref{c3}) are satisfied. Note that, our required
${\eta} _{max}$ is the maximum value of all these $\eta _{max} ({\lambda}^2)$'s. 
From our numerical results, we have plotted $\eta _{max} ({\lambda}^2)$ against
$\lambda^2$, in figure 1, which shows that the maximum value $\eta
_{max}$ of all $\eta _{max} ({\lambda}^2)$'s occurs at $\lambda = 0$ ({\it
i.e.}, at $\Lambda = 0$), the maximum value being $1/{\sqrt{3}}$. 

\section{Asymmetric optimal universal disentangling machine}

Here we apply the operations $U_x$ and $U_y$ (given in equations (\ref{U0}) and
(\ref{U1})) separately on the qubits $x$ and $y$ of the state $\left| \psi \right\rangle$ (given by equation
(\ref{twoqubit})) respectively. After taking trace over the machine states, and
using the unitarity, orthogonality and isotropy conditions ({\it i.e.},
equations (\ref{unitari}), (\ref{ortho}), (\ref{machine1}) -- (\ref{machine4}))
for both the parties $x$ and $y$, the reduced density matrix (of the two
parties $x$, $y$) becomes:
$$D_{xy} =$$
\begin{equation}
\label{general}
\left[
\begin{array}{cccc}
b_1 - c & -i \alpha \beta \Lambda_x \eta_y & -i \alpha \beta \Lambda_y \eta_x &
\alpha \beta \eta_x \eta_y\\
i \alpha \beta \Lambda_x \eta_y & b_2 + c & 0 & i \alpha \beta \Lambda_y \eta_x\\
i \alpha \beta \Lambda_y \eta_x & 0 & b_3 + c & i \alpha \beta \Lambda_x \eta_y\\
\alpha \beta \eta_x \eta_y & -i \alpha \beta \Lambda_y \eta_x & -i \alpha \beta
\Lambda_x \eta_y & b_4 - c
\end{array}
\right]_,
\end{equation}
where 
\begin{equation}
\label{bandc}
\left.
\begin{array}{lcr}
b_1 &=& \displaystyle{\frac{\left(1 - \eta_x \right) \left(1 - \eta_y
\right)}{4}~~ + ~\frac{\alpha^2 \left(\eta_x + \eta_y \right)}{2}},\\
b_2 &=& \displaystyle{\frac{\left(1 - \eta_x \right) \left(1 + \eta_y
\right)}{4}~~ + ~\frac{\alpha^2 \left(\eta_x - \eta_y \right)}{2}},\\    
b_3 &=& \displaystyle{\frac{\left(1 - \eta_x \right) \left(1 + \eta_y
\right)}{4}~~ + ~\frac{\beta^2 \left(\eta_x - \eta_y \right)}{2}},\\
b_4 &=& \displaystyle{\frac{\left(1 - \eta_x \right) \left(1 - \eta_y
\right)}{4}~~ + ~\frac{\beta^2 \left(\eta_x + \eta_y \right)}{2}},\\
c &=& 2 \alpha \beta \Lambda_x \Lambda_y.
\end{array}
\right\}
\end{equation}

Using Peres -- Horodecki theorem \cite{peres}, \cite{horodecki},
\cite{sanpera}, we see that $D_{xy}$ 
will be separable ({\it i.e.}, $\left| \psi \right\rangle$ will be disentangled) 
if the following three conditions are satisfied: 
\begin{equation}
\label{generalc1}
F_1 + 2 \alpha \beta \left\{\eta_x \eta_y \Lambda_x \Lambda_y - \alpha \beta
\left(4 \Lambda_x^2 \Lambda_y^2 + \Lambda_x^2 \eta_y^2 + \Lambda_y^2 \eta_x^2
\right)\right\} \ge 0,
\end{equation}
\begin{equation}
\label{generalc2}
F_2 + (\alpha \beta)^2 \left\{4 \alpha \beta \eta_x \eta_y \Lambda_x \Lambda_y - \left(\Lambda_x^2 \eta_y^2 + \Lambda_y^2 \eta_x^2
\right)\right\} \ge 0,
\end{equation}
$$F_3 +$$
$$\frac{\alpha^4 \beta^4}{2} \left\{4 \Lambda_x^2 \Lambda_y^2 \left(\eta_x^2 +
\eta_y^2 + 8 \Lambda_x^2 \Lambda_y^2 + 2 \eta_x^2 \eta_y^2 \right) + 16 \Lambda_x^2
\Lambda_y^2 \left(\Lambda_x^2 \eta_y^2 + \Lambda_y^2 \eta_x^2 \right)\right\}$$ 
$$+ \frac{\alpha^4 \beta^4}{2} \left(\Lambda_x^2 \eta_y^2 - \Lambda_y^2
\eta_x^2 \right) \left\{2 \left(\Lambda_x^2 \eta_y^2 - \Lambda_y^2 \eta_x^2 \right) + \eta_x^2 - \eta_y^2
\right\}$$
$$+ \frac{\alpha^3 \beta^3 \Lambda_x \Lambda_y \eta_x \eta_y}{2} \left\{\left(2
- \eta_x^2 - \eta_y^2 - 16 \Lambda_x^2 \Lambda_y^2 \right) - 4
\left(\Lambda_x^2 \eta_y^2 + \Lambda_y^2 \eta_x^2 \right)\right\}$$
$$- \frac{\alpha^2 \beta^2}{8} \left\{4 \Lambda_x^2 \Lambda_y^2 \left(1 +
\eta_x^2 + \eta_y^2 - 3 \eta_x^2 \eta_y^2 \right) + \left(\Lambda_x^2 \eta_y^2 + \Lambda_y^2 \eta_x^2 \right) \left(1 - \eta_x^2
\eta_y^2 \right) + \left(\Lambda_x^2 \eta_y^2 - \Lambda_y^2 \eta_x^2 \right) \left(\eta_x^2 - \eta_y^2 \right)\right\}$$
\begin{equation}
\label{generalc3}
- \frac{1}{8} \alpha \beta \Lambda_x \Lambda_y \eta_x \eta_y \left(1 - \eta_x^2
\right) \left(1 - \eta_y^2 \right) \ge 0,
\end{equation}
where
\begin{equation}
\label{F123}
\left.
\begin{array}{lcr}
F_1 &=& b_1 b_2 + b_1 b_3 + b_1 b_4 + b_2 b_3 + b_2 b_4 + b_3 b_4 - \alpha^2
\beta^2 \eta_x^2 \eta_y^2,\\
F_2 &=& b_1 b_2 b_3 + b_1 b_2 b_4 + b_1 b_3 b_4 + b_2 b_3 b_4 - \alpha^2
\beta^2 \eta_x^2 \eta_y^2 \left(b_1 + b_4 \right),\\ 
F_3 &=& b_1 b_2 b_3 b_4 - b_1 b_4 \alpha^2 \beta^2 \eta_x^2 \eta_y^2,\\
\end{array}
\right\}
\end{equation}
$b_1$, $b_2$, $b_3$, $b_4$ being given by equations (\ref{bandc}). Here
$\Lambda_j = \lambda_j \sqrt{(1 - \eta_j^2)/4}$, where $\lambda_j = {\rm Im}
\left\{ _j\!\left\langle M_0 | \widetilde{M_0} \right\rangle_j \right\}$ for $j
= x, y$.

Let us first consider the case when $\Lambda_x = 0$, and $\Lambda_y = 0$. In
this case, above three conditions (\ref{generalc1}) -- (\ref{generalc3}) will
be reduced to the following three conditions respectively.
\begin{equation}
\label{F1}
F_1 \ge 0,
\end{equation}
\begin{equation}
\label{F2}
F_2 \ge 0,
\end{equation}  
\begin{equation}
\label{F3}
F_3 \ge 0,
\end{equation}
where $F_i$'s are given in (\ref{F123}). Conditions (\ref{F1}) -- (\ref{F3})
will be satisfied for all $\alpha \beta \in [0, 1/2]$ if the reduction factors
$\eta_x$ and $\eta_y$ satisfy the following relation:
\begin{equation}
\label{optimalrelation}
\eta_x \eta_y \le \frac{1}{3}.
\end{equation}
Thus for given any $\eta_x$ ($\eta_y$) in $(0, 1]$, the maximum value
$\eta_{y(max)}$ ($\eta_{x(max)}$) of $\eta_y$ ($\eta_x$) is $1/3 \eta_x$
($1/3 \eta_y$).   

Next we consider the general situation. Since we are looking for universal
disentangling machine(s), thus, for given any value of the pair
$(\lambda_x, \lambda_y)$ in $[0, 1] \times [0, 1]$, and for given any value of
$\eta_x \in (0, 1]$, each of the three
conditions (\ref{generalc1}) -- (\ref{generalc3}) has to be satisfied for all
$\alpha \beta \in [0, 1/2]$, and for all values of $\eta_y \in (0, \eta_{y(max)}]$, where $0 \le \eta_{y(max)} \le 1$. 
Now we see that (i) satisfaction of the condition
(\ref{generalc1}) (universally) implies satisfaction of condition (\ref{F1})
(universally) (as the term other than $F_1$ on the left hand side of (\ref{generalc1})
becomes non-positive for $\alpha \beta = 1/2$), (ii) satisfaction of the condition
(\ref{generalc2}) (universally) implies satisfaction of condition (\ref{F2})
(universally) (as the term other than $F_2$ on the left hand side of (\ref{generalc2})
becomes non-positive for $\alpha \beta = 1/2$). But it can be shown that one
cannot find even a single value of $\alpha \beta \in (0, 1/2]$ for which the term
on the left hand side of (\ref{generalc3}), other than $F_3$, always remains 
non-positive for every choice of $\lambda_x, \lambda_y \in [0, 1]$, and
for every choice of $\eta_x, \eta_y \in [0, 1]$.{\footnote{We have verified this
numerically.}} So we have to take into account the three conditions
(\ref{generalc1}) -- (\ref{generalc3}) all together. We have obtained numerically
that for given any value of the pair
$(\lambda_x, \lambda_y)$ in $[0, 1] \times [0, 1]$, and for given any value of
$\eta_x \in (0, 1]$, the maximum value $\eta_{y(max)}$ (say) of $\eta_y \in (0,
1]$, for which all the three
conditions (\ref{generalc1}) -- (\ref{generalc3}) are satisfied for all
$\alpha \beta \in [0, 1/2]$, satisfies the relation $\eta_{y(max)} \le
1/3\eta_x$ (see figure 2 for a comparison). So the optimal disentangling machine, in the
asymmetric case, will be obtained from the case where $\Lambda_x = 0$ and
$\Lambda_y = 0$, and so the optimal curve to be satisfied by the reduction
factors $\eta_x$ and $\eta_y$, is given by the following rectangular hyperbola.
\begin{equation}
\label{rectangular}
\eta_x \eta_y = \frac{1}{3}.
\end{equation}
It is clear from condition (\ref{optimalrelation}) that (i) the maximum value
of $\eta_y$, in the totally asymmetric case ({\it i.e.}, when $\eta_x = 1$) is
$1/3$ (as shown in section 2), and (ii) the maximum value
of $\eta$, in the symmetric case ({\it i.e.}, when $\eta_x = \eta_y \equiv \eta$) is
$1/{\sqrt{3}}$ (as shown in section 3).   

Now we come to the question of disentanglement (by using local operations
only) of an arbitrary mixed state $\rho$ of the two qubits $x$ and $y$.
In its spectral representation, $\rho$ takes the following form:
\begin{equation}
\label{spectral}
\rho = \sum_i \mu_i P\left[\left| \psi_i \right\rangle\right],
\end{equation}
where $\mu_i \ge 0$, $\sum_i \mu_i = 1$, and $P\left[\left| \psi_i
\right\rangle\right]$ is the projection operator on the (normalized) pure state
$\left| \psi_i \right\rangle$ of the two qubits $x$ and $y$. One can 
express $\left| \psi_i \right\rangle$, in its Schmidt form, as 
\begin{equation}
\label{psii}
\left| \psi_i \right\rangle = a_i \left| 0_i 0_i \right\rangle_{xy} + b_i
\left| 1_i 1_i \right\rangle_{xy},
\end{equation}
where $a_i$, $b_i$ are non-negative numbers with $a_i^2 + b_i^2 = 1$, and
$\left| 0_i \right\rangle_j$, $\left| 1_i \right\rangle_j$ are two orthonormal
states in the Hilbert space of the qubit $j$ (for $j = x, y$). As discussed
earlier, each of the states $\left| \psi_i \right\rangle$ will be disentangled
by using local operations, with some common values of the reduction factors
$\eta_x$ and $\eta_y$. Let ${\rho}_{j(\psi_i)}^{bd}$ and ${\rho}_j^{bd}$ be the
single particle reduced density matrices of $\left| \psi_i \right\rangle$ and
$\rho$ respectively, corresponding to the qubit $j$ ($j = x, y$), before 
disentanglement; and let ${\rho}_{j(\psi_i)}^{ad}$ and ${\rho}_j^{ad}$ be the
single particle reduced density matrices of $\left| \psi_i \right\rangle$ and
$\rho$ respectively, corresponding to the qubit $j$ ($j = x, y$), after
disentanglement. If $\eta_j^{\prime}$ is the reduction factor associated with
the disentanglement of $\rho$, corresponding to the qubit $j$ ($j = x, y$), we then have
\begin{equation}
\label{relation1}
{\rho}_j^{ad} \equiv \eta_j^{\prime} {\rho}_j^{bd} + \frac{1 - \eta_j^{\prime}}{2} I
= \eta_j^{\prime} \sum_i \mu_i {\rho}_{j(\psi_i)}^{bd} + \frac{1 - \eta_j^{\prime}}{2} I,
\end{equation}
or
\begin{equation}
\label{relation2}
\sum_i \mu_i {\rho}_{j(\psi_i)}^{ad} \equiv \sum_i \mu_i \left\{\eta_j
{\rho}_{j(\psi_i)}^{bd} + \frac{1 - \eta_j}{2} I \right\} = \eta_j^{\prime} \sum_i \mu_i {\rho}_{j(\psi_i)}^{bd} + \frac{1 - \eta_j^{\prime}}{2} I.
\end{equation}
Thus we get from equation (\ref{relation2}) that $\eta_j^{\prime} = \eta_j$ for
$j = x, y$.  
        
\section{Discussion} 

Since an ideal universal disentangling machine does not exist, we have explored 
how well a universal disentangling machine can be, using most general type of
local operations. We have shown that in the case of optimal universal disentangling
machines, reduction factors lie on a rectangular hyperbola. Here one should
mention that in the case of
optimal universal clonning machines, the relation between the reduction factors
follows directly from no--signalling phenomenon \cite{ghosh}. So one may
look for the physical phenomenon (or phenomena), from which above--mentioned rectangular hyperbola would follow
directly. This is still an unresolved problem.   

We now address the issue of obtaining a better disentangling machine, if
possible, using non-local operations. 

For all (non-unitary) local operations, entanglement decreases, whereas, for
non-local operations, it may increase, decrease, or remain same. In the
disentanglement process described here, a separable state remains separable
without any further constraints (other than isotropy condition) on the
machine,  but even in order to keep  a separable state separable 
under non-local 
operations,
some constraints other than isotropy condition have to be imposed on the 
machine, which seems to decrease the reduction factor. But if one redefine the
notion of disentanglement as a process which also keeps every pure bipartite product
state, a product of two (single particle) density matrices of the two
particles, then the allowed class of disentangling machines will comprise of
only local operations.
 
In this regard, for intuitive understanding, we
point out that local cloning machine, with the blank copy, functions as universal
disentangling machine  which can be made optimal. Now if we use optimal local cloning
machine, which produces three copies instead of two, then the cloning machine
along with the two blanck copies acts as a universal disentangling machine with
reduction factor being 5/9 \cite{som1}. A non-local cloning machine \cite{buz} along with six blanck
copies, which produces seven copies of the bipartite states, acts as universal
disentangling machine with reduction factor being 11/35 \cite{som1}, 
which is much less than the former case. Recently Mor and Terno \cite{mt}
obtained a set of bipartite entangled states, which can be perfectly
disentangled, provided the reduced density
matrices of one of the parties commute. But
this is achieved using local operation, namely local broadcasting.

In conclusion, we have obtained the optimal disentangling machine exploiting
the most general local operations, and discussed that there can not be
any non-local operation which may give a better one.

\vspace{0.4cm}
{\noindent {{\bf Acknowledgement}: The authors thankfully acknowledge the
sugessions and comments of the anonymous refree for revision of the earlier
version of the manuscript.}}

\vspace{0.7cm}
{\noindent {{\bf FIGURE CAPTION} : Figure 1 describes the decreasing nature of
${\eta} _{max} ({\lambda}^2)$ with the increament of ${\lambda}^2$. This graph
shows that the maximum value ${\eta} _{max}$ of ${\eta} _{max} ({\lambda}^2)$ occurs at
${\lambda}^2 = 0$, and (so) ${\eta} _{max}$ is equal to $1/{\sqrt{3}}$.}}      

\vspace{0.4cm}
{\noindent {{\bf FIGURE CAPTION} : Figure 2 shows the optimal curve $\eta_x
\eta_y = 1/3$, which corresponds to the case $\lambda_x = 0$, $\lambda_y =
0$ (represented in the figure by the continuous line). Also, for a comparison, the optimal curves corresponding to the cases
$(\lambda_x, \lambda_y) = (0.2, -0.2)$ (represented in the figure by the broken
line of the type `- -~~ - -'), $(\lambda_x, \lambda_y) = (0.5, -0.5)$ (represented in the figure by the broken
line of the type `--~~ --'), $(\lambda_x, \lambda_y) = (0.9, 0.1)$ (represented in the figure by the broken
line of the type `- - -~~ - - -') are given here.}}  

\end{document}